# Recapitulation of Web Services based on Tree Structure

Ashok kumar.P.S
Research Scholar,
Anna University, Coimbatore,

G.Mahadevan,
Prof & Head
AMCE, Bengaluru

Gopal Krishna.C
Asst. Prof, Dept. of CSE
AIT, Chikmagaluru

## ABSTRACT
The Recapitulation of Web service is an approach for the effective integration of distributed, heterogeneous and autonomous applications to build more Structured and value added services. Web services selection algorithms are required to find and select the best services. A QoS is a benchmark to select the best service for the task of composition. The importance of the web services selection algorithm is to maximize the QoS of the web services recapitulation using complex service provider's (CSP) QoS requirement parameters. This paper have discussed about different Web service Service-offers. We have classified the CSP's requirements defined on the QoS and Service-offers based on its structure. We have proposed a tree structure to represent the CSP's requirement to be defined based on the multiple QoS properties.

## 1. INTRODUCTION
A Web service is an API that describes a collection of operations that are accessible through Internet based on standardized XML messaging. Web services can be published, found and used across the Internet using SOAP, WSDL and UDDI standards. The Web service architecture is based on the interactions between service requester, service registry and service provider, where the interactions between the services involves publish, finding and bind operations [1]. QoS (Quality of Service) is a combination of Web service qualities, and it evaluate of how well the Web service convey the information to requester.

Web services are loosely coupled in nature. The travel agent doesn't need to have prior agreements with service providers or credit card companies [3]. This allows the travel agent to access more services, offer more options to their customers, i.e. the credit card companies offer their services broadly to make their customers happy, as the service providers offer their services broadly and easily they are generating more business for themselves. QoS can be used for selection and ranking of the Web services by extending standard service oriented architecture (SOA) [1]. In this paper, we have defined various Service-offers and proposed a tree structure to represent the CSP's requirements limited in the QoS and Service-offers [4].

**Case study 1:** Travel Reservation Scenario Application;

The travelers normally prefer reservation for his/her distant travel location through travel agent. The main objective of a traveler is to get the best combination of services like, quality, price and valid offers which satisfy his/her needs. On the other hand, travel agent tries to satisfy the customer's needs and mint money by charging extra fees like service charge for each trip.

In this Travel Reservation Scenario Application, travel agent is a service Intermediator, who has to find the best services for the individual tour package based on the traveler's demands on the service quality and offers. Figure 1 shows the Travel Reservation scenario Application. The travel agent uses the Web service system to find and integrate different services that are provided by the different travel service providers. The travel agent service can publish the specific service information into service registry, but QoS of the service is permitted to create a document either in digital or hard copies, in whole or part of a traveler.

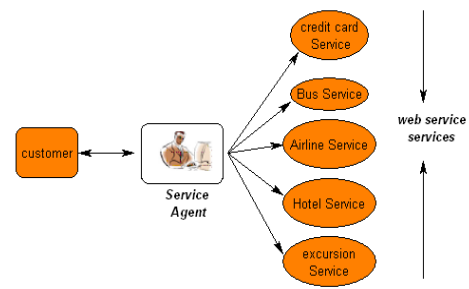

**Figure 1: Travel Reservation Scenario**

To republish, or to post on servers, agent requires prior specific permission or a traveler has to pay specified trip fee, depending on the QoS of constituent services. With respect to the above mentioned example, the customer's Travel Reservation Scenario application consist following activities:

(a) Book a Air ticket from Bangalore to Malaysia.
(b) Book a single AC Room in Malaysia star Hotel for 02 days.
(c) Book a Taxi in Malaysia from airport to Hotel.
(d) Book a ticket for Malaysia city tour.
(e) Booking for the dinner at Malaysia Beach Hotel.

The customer's QoS and business offer requirements are as follows –
- ❖ The price should be minimum.
- ❖ The most esteemed service offers with good discount.

When travel agents get such requirements from a traveler, the agent has to find the service that satisfies all requirements of a traveler. Usually a Travel agent is interested in reliable travel service provider to improve the reliability of a travel composite service [1] [5].

The Composite or Complex service provider (CSP) defines the requirements to travel agents on the multiple QoS properties and Service-offers involving AND/OR operators, it is very tedious work to find the best Web services for the individual task of the recapitulation. This paper addresses few issues related to service selection in recapitulation.
The contributions of this paper are:

• Extension of business QoS model by introducing the Concept of Service-offers.
• Definition of various Service-offers of Web services.
• Categorization of CSP's requirements
• Tree model to represent CSP's requirements based on the multiple QoS properties and business offers.

This paper is organized as;
Section 2 specifies and categorizes the various Service-offers, Section 3 defines the QoS Model of service offers, Section 4 defines the classification and tree structure of CSP's requirements and Section 5 depicts the conclusion and future work.

## 2. SERVICE-OFFER MODEL
In order to attract the customers, the service providers normally advertise a lot, as well as display with attractive offers to improve their business with huge profit [10]. In this section, we define Service-offer vocabulary from the requester's point of view. This paper define the Service-offers as a reduction in the service price or providing the services as they advertise or offering an article as a gift for the service consumption, etc.. [1][7]. This paper broadly categorizes various web services Service-offers like,

   i.   *Value based offers,*
   ii.  *Gift Based Service-offers*
   iii. *Lucky Service-offers*
   iv.  *Agreement Service-offers*

## 2.1 Value Based Service-Offers
Value based offers are nothing but an unconditional discounts or cash gifts for each and every service activities. These offers are given to the customers without any preconditioned terms & conditions [10]. We identified *the* value based offer like,
   1. *Discount offer.*

### 2.1.1 Discount Offer (DO).
Discount offer consists of discount or concession in service price for every service activities, which is expressed in terms of percentage. For example tour package provider may offer 15% discount in early bird bookings.

## 2.2 Gift Based Service-offers
Gift based service-offers normally consists of gift items or free services for the service consumption. For example, the air ticket reservation provider may offer a gift hamper or a free service [10]. We define *two* types of commodity based Service-offers.

### 2.2.1 Article Offer (AO).
In an article offer, the service provider may offer an item as a gift for the service activity. For example, the reservation service provider may offer a gift article for the confirmed reservation.

### 2.2.2 Service-offer (SO).
The *Service* provider will offer the free of cost service as a gift for the service activity. For example, the Reservation service provider may offer one free reservation after one paid reservation.

## 2.3 Lucky Service-offers
The offers that are probabilistic in nature are called as Lucky Service-offers. We define *one* Lucky Service-offer as,

### 2.3.1 Lucky Coupon Offer (LCO)
A lucky coupon offer is a chance (probabilistic) based offer, where the lucky coupon of specified amount is given on the service requester without any preconditions. For example, on reserving two air tickets, the service provider may offer a lucky coupon worth Rs.500 or give fancy article.

## 2.4 Agreement Service-offers
The agreement offers are defined on the precondition described by the service provider. For example, the air ticket reservation service provider may offer a discount of 10% on booking air ticket. We define *two* types of agreement Service-offers.

### 2.4.1 Conditional Service-offer (CSO)
A conditional Service-offer is a conditional offer, which provides a free service based on the precondition defined on the number of service executions. For example, a reservation service may advertise that, 'reserve an Accommodation and get Food free in our hotel '.

### 2.4.1 Conditional Discount Offer (CDO)
A conditional discount offer is a conditional offer, which offers a reduction in service price under the precondition defined on the number of service executions. For example, a reservation service may advertise 10% reduction for a reservation after two tickets are confirmed reservations.

## 3. QOS MODEL
QoS model of Web service includes the Service-offers with in the business specific QoS category. Figure 2 shows the major QOS model of Business Specific Offer categories. The service provider can advertise multiple offers which may belong to the same kind or different kind [7] [8]. For example, the travel reservation service provider may advertise offers like, 'confirm four reservations, one reservation and one lucky coupon is free'; these Service-offers are recorded in a separate *Service-offer table.* Usually the Service-offer table contains mandatory information like, Types of Service-offer, information of the offer and service-keywords of the specific advertised Service-offers are specified in Table 1.

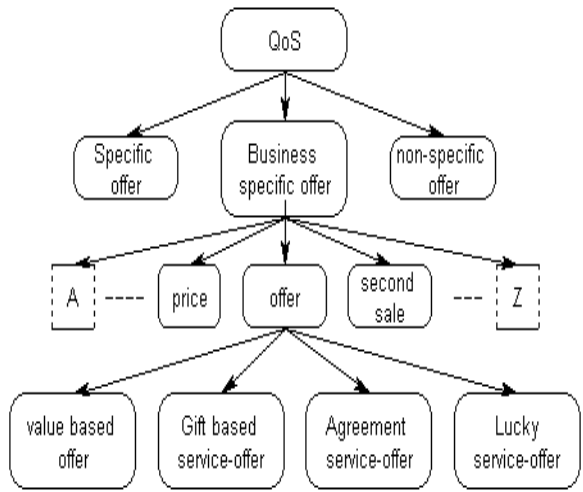

**Figure 2: QOS model of Business Specific Offer**

The QoS is most important for Web service publishing and selection of service activities [2]. The parameters in a Service-offer are, *price* is a real number which indicates the currency, *percentage* refers to a number in the range, *period* refers to time period in hours or days *and frequency* refers to number of service executions (always greater than one). The Service-offer listings which are defined in this model have to be used by all Web service providers and service requesters for the Service-offer.

**Table 1. Service Qualities & Information**

| Service-offer Types | Offer Information |
|---|---|
| Cash offer (CO) | Price |
| Discount offer (DO) | Percentage (p) |
| Article offer (AO) | Price |
| Service-offer (SO) | Price |
| Lucky coupon offer(LCO) | price, period |
| Conditional Service-offer(CSO) | Frequency |
| Conditional discount offer (CDO) | Frequency(F), Percentage |

## 4. REQUIREMENTS OF WEB SERVICES RECAPITULATION

*Recapitulation requirements* are the requirements which are defined based on the QoS and Service-offered by the CSP [4]. The recapitulation requirements different for each and every service providers, which are dependent on the customers or requester's basic needs or service provider's competition of service or target in achieving the profit. The CSP also gives preference or weight age to each service requirement. For example, the CSP may compose the services, which are low cost and fast response or low cost with high reputation, security and discount offer etc. consider the travel package scenario; here the requirements on the service price and service offers that are necessary to execute the complex service in a profitable manner. Thus CSP can impose different requirements on several QoS properties and/or Service-offers with varied preferences.

For example, a Recapitulation requirement of a customer is requested for a Service-offer to choose a Flight as well as Hotel Accommodation in Malaysia, through service Agent. The recapitulation requirements service offer steps are given below,

1. The travel agent service finds a list of airlines.
2. The customer communicates his/her choice for the flight.
3. The airline returns a confirmation number with an expiry date.
4. The travel agent service requests the chosen airline to put the flight on hold:
    i. The travel agent service requests a description of how to put a seat on hold to the airline service.
    ii. The travel agent service sends the request based on customer's priority.
5. Identification of Accommodation in the Hotel by Travel Agent.
    i. The travel agent communicates with the Hotel manager for the description of service found.
    ii. He requests the accommodation options for the period.
6. The travel agent service looks for the payment services available, and builds a list of options for the customer.
7. The travel agent service delivers the results of the queries to the customer to choose their best option, along with the payment options that offered.

### 4.1 Recapitulation Requirement Types
Recapitulation requirements are different for service providers and service requester, for their Reputation and security. we classify complex service provider's (CSP) requirements based on the requirement structure as;
- ❖ Simple requirements
- ❖ Complex requirements.

A *simple requirement* is defined based on a single QoS property or a single Service-offer. For example, the CSP might say, 'they provide minimum cost of the Hotel Accommodation service '. If the simple requirement is defined on the QoS property, then it is called as a *Simple Quality Requirement*.

A *complex requirement* is composed of two or more simple requirements using recapitulation operators AND & OR. For example, the CSP might say, 'they provide minimum cost and more discount in their services '. The CSP can enforce either a simple or complex requirement with varied preferences for the selection to build a desired quality complex service.

### 4.2 Recapitulation Requirement Modeling

The recapitulation requirements are to be fulfilled based on the QoS properties and/or Service-offers with reference to the simple and/or complex web service requirements [3]. In this paper, we propose a tree structure called *Recapitulation Requirement Tree, which* represents the complex service provider's (CSP) recapitulation requirements.

### 4.2.1 Recapitulation Requirement Tree (RRT)

A recapitulation requirement tree is a AND-OR tree, with Weighted node [1] whose leaf node contains the QoS property or Service-offer. For Example: the label $A_{pq}$, on the edge between any two nodes 'p' and 'q', represents the preference for the sub-tree rooted at the node 'q' while traversing from root to leaf i.e. the edge label represents the preference to either simple or complex requirement [6].

For Example: if we consider the travel agent's requirements in order to increase the profit in the travel package booking, he has to follow:

(1) Discount offer (DO) or Service offer (SO)
(1) Popular (PO) or Lucky coupon offer (LCO)

The CSP expects both the requirements to be equally satisfied for the recapitulation with preferred weights 0.4, 0.6, 0.7 and 0.3 to Discount offer, Service Free, Lucky coupon offer respectively. This recapitulation requirement can be represented as the RRT in Figure 3.

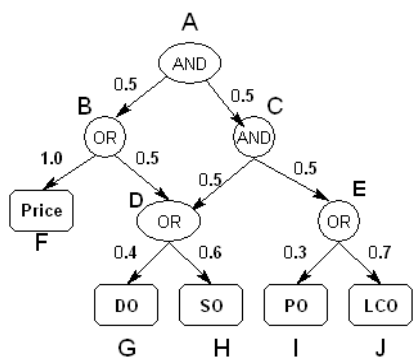

Figure 3: Recapitulation Requirement Tree

The node J represents the simple offer requirement and the node H represents the simple quality requirement. The sub-tree at node D represents the complex requirement and the weight (0.5) on the edge (B, C) represent the preference for the complex requirement defined at node B.

### 4.2.2 Estimation of the Service-offers

The customer service provider publishes a complex Web service Service-offers. The discovery mechanism in a Web service normally discovers in a collection of complex Web services with different Service-offers and QoS parameters. In order to find the Web services based on the Service-offers, we are representing a computational parameter called *Profit (Pr)*, which is calculated as the ratio of the profit amount to the requester payable amount (Amount). The Profit "Pr" is computed based on the types of Service-offer. Table 2 represents how to compute the profit for every service-offer.

**Table 2 – Profit computation for Service-offer**

| *Service-offer* | *Profit (Pr)* |
|---|---|
| Cash offer | $Pr = \dfrac{price}{offer\,price}$ |
| Discount offer | $Pr = \dfrac{\% \times offer\,price}{100}$ |
| Article offer | $Pr = \dfrac{price}{Quantity \times Item\,value}$ |
| Services offer | $Pr = \dfrac{price}{offer\,price}$ |
| Lucky coupon | $Pr = \dfrac{price}{offer\,price}$ |
| Conditional Service-offer | $Pr = \dfrac{price}{Frequency \times offer\,price}$ |
| Conditional discount offer | $Pr = \dfrac{\% \times offer\,price}{Frequency \times 100}$ |

### 4.2.2 Web services Selection Mechanism

In Web services recapitulation, the selection mechanism accepts the Recapitulation Requirement Tree (RRT) and requester Web services as an input and assigns the best Web service for that particular task. Here Web services Selection mechanism in RRT traverses in a bottom-up approach [1].

At the bottom most leaf nodes the selection mechanism performs *two* actions, like:
  (1) *Scaling*
  (2) *Ranking*.

In the *scaling phase*, the Web service selects a simple offer requirement. In simple offer requirement the price (P) of a Web service is calculated. Now the P/QoS values of the selected Web services are normalized using min-max normalization technique [7].

In *ranking phase*, the normalized values are multiplied with the weight to get new scores for Web services [9].

At the internal node, the selection mechanism performs *two* actions:
  (1) *Filtering*
  (2) *Ranking*

*Filtering & Ranking* selection mechanism are dependent on the types of internal node *AND/OR* operations.

In *filtering phase*, if the node is *AND*, then the Web services present in *entire* child nodes are get selected.
If the node is *OR* then the *distinct* Web services in the descending order of their scores are selected from its child nodes.

In *ranking phase*, if the node is *AND*, then the score of selected Web service is computed. When the sum of the scores of selected Web service at its child nodes are multiplied, it results with the weight of sub-tree rooted at *AND* node [9].

If the node is *OR* then the score of selected Web service is multiplied with the weight of sub-tree rooted at *OR* node. After ranking the Web services at the root node, they are sorted in the descending order of their scores, and the first Web service becomes the *best choice* for the task satisfying the CSP's recapitulation requirements.

## 5. CONCLUSION

In Business offering Service-offers, *Value based offers*, *Gift Based Service-offers, Agreement Service-offers* and *Lucky Service-offers* are most important for Web service selection where the complex service provider (CSP) looks for profit out in the business. In this paper, we have defined various Service-offers for the Web services selection and estimated the profit for service requester, as well as service agents and service providers for the business. This paper intends a tree structure called *Recapitulation Requirement Tree (RRT)* to represent the multiple QoS properties and Service-offers involving AND/OR operators with varied preferences. Based on QoS confined complex service selection algorithms, we can discover the near-optimal complex web services efficiently.

In our future work, we plan to add a novel trust evaluation approach that aggregates the ratings from other customers or requesting client's to the service agents prior subjective belief about the trusted service.